\documentclass[amsart]{article}
\usepackage[margin=1.5 in]{geometry}
\usepackage{authblk}
\usepackage{graphics}
\usepackage{caption}
\usepackage{mathtools}
\usepackage{amsmath}
\usepackage{mathrsfs}
\usepackage{graphicx}
\usepackage{wrapfig}
\usepackage{epstopdf}
\usepackage{enumitem}
\usepackage{lipsum}
\usepackage{placeins}
\title{Thermodynamics of fluctuations in small systems interacting with the environment}
\date{}
\author[1]{Salvatore Calabrese}
\author[2]{Lamberto Rondoni}
\author[1*]{Amilcare Porporato}

\affil[1]{Department of Civil and Environmental Engineering and Princeton Environmental Institute, Princeton University, Princeton, NJ, USA.}
\affil[2]{Dipartimento di Scienze Matematiche, Politecnico di Torino and INFN, Sezione di Torino, Torino, Italy}
\affil[*]{Corresponding author: Amilcare Porporato, aporpora@princeton.edu}

\begin{document}
	\maketitle

A few decades after Hill's work on nano-thermodynamics, the development of a thermodynamic framework, to account consistently for the fluctuations of small systems due to their interactions with the surrounding environment, is still underway. Here we discuss how, in a small system, the interaction energy with the environment may be described through a conjugate pair of intensive and extensive variables, giving rise to a Gibbs thermodynamics with an additional thermodynamic degree of freedom. The relevant thermodynamic potentials help describe the equilibrium conditions and the material properties that measure the susceptibility of the system to the interaction with the environment. The resulting generalized ensembles, which describe the non-negligible, small system fluctuations, are shown to be equivalent. Away from the average thermodynamic state, the availability describes the distance between system and the environment, in terms of the maximum work extractable from the fluctuation, and a proper definition of entropy extends the thermodynamics to the generic fluctuating state. From the latter state, entropic forces arise to restore the average thermodynamic state. Our framework unifies and extends the ensemble thermodynamics by Hill and the recent advances in statistical mechanics under strong coupling and reduces to classical macroscopic thermodynamics when the system is large. The particular case of a small ideal gas system is discussed in detail and the example of a single particle immersed in a bath is revisited in the light of the new formalism.


\section{Introduction}

The possibility of observing and manipulating systems at the nanoscale is providing a unique opportunity to investigate the origin of non-equilibrium behavior of biological systems, from the activity of individual cells to the complex molecular structures and beyond \cite{bustamante2004mechanical,nelson2004biological,kinbara2005toward,shipley2010plant,abendroth2015controlling}. Spurred by these recent theoretical advances and the novel manipulation techniques in nano-systems (e.g., optical tweezers), the thermodynamics of small systems has received renewed attention in the past two decades  \cite{hummer2001free,collin2005verification,ciliberto2010fluctuations,an2015experimental}. Differently from macroscopic systems treated by classical thermodynamics, in small systems fluctuations are typically non negligible; moreover, because the interaction energy between the system and the environment is typically of the same order of magnitude as the internal energy of the system, the latter is not additive.

Stochastic thermodynamics has aimed to provide a thermodynamic interpretation to the random dynamics of particles trajectories. While early studies focused on Brownian particles and noninteracting systems \cite{sekimoto1998langevin,seifert2012stochastic,zhang2012stochastic,van2015ensemble}, recent work extended the framework to systems in strong coupling with the environment \cite{esposito2010entropy,seifert2016strongcoupling,talkner2016open,jarzynski2017stochastic,miller2017entropy}, in both equilibrium and nonequilibrium conditions. In this line of research, however, the definition of the thermodynamic quantities for a small system is not directly consistent with that for macroscopic systems \cite{sekimoto1998langevin}. This has motivated an interesting debate \cite{vilar2008failure,peliti2008comment,horowitz2008comment,hilbert2014thermodynamic,goldstein2017statistical,jarzynski2017stochastic}, calling for a consistent link between stochastic and macroscopic thermodynamics \cite{jarzynski2017stochastic}.

Earlier, a thermodynamic perspective on small systems was pioneered by Hill \cite{hill1963thermodynamics,hill2001different}. Based on the realization that small systems have an additional thermodynamic degree of freedom, he extended classical thermodynamics by considering multiple replicas of the given system. Because the replicas are independent realization of the given small system, the internal energy of the ensemble of replicas is additive and their thermodynamics is extensive. While Hill's approach has been successfully used \cite{chamberlin2000mean,schnell2011thermodynamics,schnell2012thermodynamics,chamberlin2014big,latella2017long}, Hill's nano-thermodynamics is extensive only for the ensemble of replicas, whereas it remains non-extensive for the single small system. Consequently, while the Gibbs equation still holds, the Euler relation is corrected with an additional \text{\it ad hoc} potential. This makes it difficult to fully connect Hill's and Gibbs' thermodynamics \cite{qian2012hill,bedeaux2018hill}.

Stochastic thermodynamics has largely progressed independently of Hill's thermodynamics. The objective of this contribution is thus to present a small system thermodynamics, {\it  \`{a} la Gibbs}, which encompasses both Hill's thermodynamics and the recent advances in statistical mechanics under strong coupling. Towards this goal, the first logical step is to describe the average thermodynamic state by means of a fundamental equation that includes the additional degree of freedom in the list of extensive quantities (in the sense of Gibbs \cite{gibbs1928collected}, p. 85; see also \cite{callen2006thermodynamics}). Gibbs and Euler equations for the small system are then derived along with novel and more convenient potentials (by Legendre transformation). Among these, Hill's subdivision potential, appearing as the product of conjugate variables associated to the interaction between system and environment, acquires a more physical meaning. The material properties for small systems are also generalized, including new ones that describe the effect of the interaction with the environment. Besides a proper definition of the interaction potential for a correct interplay between energy and entropy, small systems also require a proper definition of volume and its fluctuations. We do not address this problem here but refer to the previous literature (e.g., \cite{corti1998deriving,hunjan2010voronoi,uline2013molecular}) to guarantee consistency with the macroscopic definition of volume. 


Small systems are most likely found in a fluctuating condition about what can be seen as an average thermodynamic state. Here, we focus on fluctuations due to the interaction with the environment and not of small systems in isolation, which may not settle in a "thermodynamic" state \cite{campa2009statistical}. We therefore derive a generalization of the thermodynamic theory of fluctuations, which describes the energetics of the fluctuating state and provides the solvated ensemble \cite{jarzynski2017stochastic} and Hill's unconstrained ensemble \cite{hill2001different} as particular cases. The availability \cite{kestin1979course} quantifies the distance of the fluctuating state from the average one in terms of maximum work extractable from the fluctuation (or minimum work required to induce it) and a proper definition of fluctuating entropy satisfies Euler and Gibbs equations, thus providing a consistent thermodynamics of the fluctuating state. Due to the concavity of the entropy function, any couple of thermodynamic ensembles for small systems are equivalent  within this framework. 

We will further discuss the case of a small system of ideal gas particles. Its peculiarity stems from the fact that the interaction between the particle within the system, and hence between the system and the environment, is negligible. As a consequence, classical macroscopic thermodynamics applies without additional degree of freedom, even if the system consists of a single ideal gas particle. To illustrate the details of the theory, we discuss the example of a single particle immersed in a heat bath and subject to an external force. We conclude by summarizing the results and by discussing the analogy between small systems and systems with long-range interactions. Our description is limited to spontaneous fluctuations around equilibrium; the role of forces driving away from equilibrium and the effect of transitions between equilibrium states will be explored in other contributions.     

\section{Thermodynamics of the average state}
\label{sec:smallsyst} 
Consider a small system, interacting with the surrounding environment, having an average state given by an internal energy $U$, volume $V$, number of moles $N$, and entropy $S$. A crucial point in these systems is that in the Gibbs equation the interaction between system and environment in general cannot be neglected \cite{gibbs1928collected,callen2006thermodynamics}. The latter is expressed by the work, $f dL$, that is exchanged between them,
\begin{equation}
\label{eq:gibbseq}
dU=TdS-pdV+\mu dN+f dL,
\end{equation}
where $f$ is the intensive quantity (e.g., a force) giving rise to the interaction between system and environment and $L$ is the corresponding extensive quantity (e.g., the coordinate on which the force operates). 

The effect of the interaction with the environment is to create an internal inhomogeneity in the system. Such an interaction brings `order' (i.e., structure) in the system, hence reducing its entropy. An interesting example from biology is the folding of a protein, which emerges from the hydrophobic interaction of the protein constituents (amino acids) and the surrounding medium \cite{nelson2004biological,dobson2003protein}. Here $f$ is the strength of hydrophobic interaction, while $L$ measures the folding. As the interaction with the environment reduces (e.g., by weakening the hydrophobic interaction with low temperatures \cite{kauzmann1959some}), the disorder in the protein configuration increases until the protein unfolds (protein denaturation). This effect is analogous to one observed in a system (even macroscopic) immersed in an external field, whose thermodynamics is also described by the Gibbs equation (\ref{eq:gibbseq}) \cite{garrod1995statistical,callen2006thermodynamics,calabrese2019origin}. For example, under the influence of a gravitational field the thermodynamic system is subject to a force that shifts its center of gravity. The infinitesimal work is $f dL=\rho N g dr$, where $\rho$ is the molar density, $g$ is the intensity of the gravitational field, and $dr$ is the infinitesimal displacement. 

A clear advantage of equation (\ref{eq:gibbseq}) is that, unlike Hill's nano-thermodynamics \cite{hill1963thermodynamics,hill2001different}, $U$ remains a linear homogeneous function of its arguments, so that the Euler relation is
\begin{equation}
\label{eq:euler}
U=TS-pV+\mu N+fL.
\end{equation}
In contrast, Hill's thermodynamics is derived at the ensemble level, whereby a Gibbs equation is formulated for a fictitious collection of a large number of replicas of the small system under study. As a result, only the total energy of the ensemble of replicas, but not the energy of the single small system, is a linear homogeneous function with all the properties of classical thermodynamics. For Hill (see Eq. (6) in \cite{hill2001different}), $f L$ is a subdivision potential, $E$, which  cannot be altered, i.e.,
\begin{equation}
\label{eq:euler2}
U=TS-pV+\mu N+E,
\end{equation}
so that its total differential is
\begin{equation}
\label{eq:gibbseq2}
dU=TdS-pdV+\mu dN,
\end{equation}
rather than (\ref{eq:gibbseq}). Hence the present framework is a generalization of Hill's approach, the latter being limited to the case in which the interaction between system and environment cannot be manipulated (i.e., constant $L$). Only under the latter condition is the thermodynamics described by the Euler relation (\ref{eq:euler}) (or (\ref{eq:euler2})) and by the Gibbs equation (\ref{eq:gibbseq2}). When the size of the system increases, the internal energy grows, as opposed to the interaction energy, which remains localized at the system/environment interface. As the thermodynamic limit is approached, the interaction energy can be neglected (i.e., $fdL \to 0$) and the formalism reduces to the classical thermodynamics of macroscopic systems.





\subsection{Extended free energy}

\label{sec:freeen}
The thermodynamics for small systems developed above has an additional degree of freedom, $L$, embedded in the fundamental equation in energy representation, $U=U(S,V,N,L)$. Such extension of the thermodynamic structure is transfered to all the other thermodynamic potentials by Legendre transformation. Consider, for example, a small system immersed in a much larger environment at constant $T$ and $p$ conditions (e.g., molecular motors in cells), in which case the Gibbs free energy representation is typically the most convenient. By Legendre transform of $U$ with respect to $S$ and $V$, the differential of the Gibbs free energy, $G(T,p,N,L)=U+pV-TS$, is
\begin{equation}
dG=-SdT+Vdp+\mu dN+f dL,
\end{equation}
which at constant $T$, $p$, and $f$ reduces to $dG=\mu dN+f dL$. For a multicomponent system, we thus have $dG=\sum_i \mu_i dN_i+f dL$, so that, even in chemical equilibrium for which $\sum_i \mu_i dN_i=0$, the Gibbs free energy is not at the minimum, since
\begin{equation}
\label{eq:gibbsfe}
dG=f dL.
\end{equation}
Equation (\ref{eq:gibbsfe}) is a result of the fact that the interaction with the reservoir induces an inhomogeneity within the system, which at constant $T$ and $p$ corresponds to further work that can be extracted from the system. 

An alternative thermodynamic representation derives from the Legendre transform of $U$ with respect to $S$, $V$, and $L$. This yields the `extended' Gibbs free energy, $G^e(T,p,N,L)=U+pV+f L-TS$, the differential of which reads
\begin{equation}
\label{eq:gibbsfeg}
dG^e=-SdT+Vdp+Ldf+\mu dN.
\end{equation}
At constant $T$, $p$, and $f$, (\ref{eq:gibbsfeg}) reduces to the well known relation $dG^e=\mu dN$, and in chemical equilibrium
\begin{equation}
dG^e=0.
\end{equation}
Therefore, for a small system it is the extended Gibbs free energy $G^e$ which has a minimum at equilibrium, while $G$ is not at a minimum and $dG$ measures the available energy (i.e., maximum work). 

The previous considerations provide another angle to see how our results, based on the Gibbs equation (\ref{eq:gibbseq}), generalize Hill's nano-thermodynamics. The differential of the term $fL$, which corresponds to the subdivision potential $E$, can be obtained by differentiating equation (\ref{eq:euler2}), and canceling the terms that satisfy (\ref{eq:gibbseq})
\begin{eqnarray}
\label{eq:gibbduhext}
Ldf=-SdT+Vdp+Nd\mu,
\end{eqnarray}
which corresponds to the Legendre transform of $U$ with respect to $S$, $V$, and $N$, and shows that $T$, $p$ and $\mu$ can be treated as independent variables, contrarily to macroscopic thermodynamics where they are related by the Gibbs-Duhem relation. By differentiating $E$ and combining with equation (\ref{eq:gibbseq2}), one can also see that $dE=-SdT+Vdp+Nd\mu$, but this does not correspond to a Legendre transform of $U$ in Hill's nano-thermodynamics.

\subsection{Material properties}
The above considerations have also important implications for the proper definition of the material properties of small systems, which are found as the elements of the Hessian matrix of $G^e$. The diagonal entries give
\begin{equation}
\label{eq:matprop}
c_{p,f,N}=T\left.\frac{\partial^2 G^e}{\partial T^2}\right|_{p,f,N},  \quad \quad K_{T,f,N}=\left.\frac{\partial^2 G^e}{\partial p^2}\right|_{T,f,N}, \quad   S_{T,p,N}=\left.\frac{\partial^2 G^e}{\partial f^2}\right|_{T,p,N}, \quad \text{and} \quad \Gamma_{T,p,N}=\left.\frac{\partial^2 G^e}{\partial N^2}\right|_{T,p,f},
\end{equation}
where: $c_{p,f,N}$ is the isobaric heat capacity at constant $f$ and $L$; $K_{T,f,N}$ is the isothermal compressibility at constant $f$ and $L$; $S_{T,p,N}$ is the isothermal susceptibility to the interaction at constant $p$ and $N$, which is a new material property measuring the capability of the environment to affect the system (see the example in Section \ref{sec:app}); and $\Gamma_{T,p,f}$ is the isothermal change in chemical potential for changes in $N$ at constant $p$ and $f$. 

From (\ref{eq:matprop}) it follows that all the material properties depend on $f$, that is on the specific environment with which the system is interacting. Thus, from a practical point of view, for systems interacting with the environment, the measurement of a material property requires either negligible or constant $f$, otherwise inconsistent estimates are obtained. This may not always be possible in small systems. Well-known situations of this kind include the anomalous transport of matter and of energy in highly confining environments. In these cases, the behavior does not correspond to that of usual aggregation states of matter, as the physical laws depend in various ways on the boundaries (the environment) \cite{kunz1994multiple}.

Proceeding with the mixed derivatives, the field-thermal coupling is described by
\begin{equation}
\label{eq:thefieldMP}
\gamma_{p,N}=\left.\frac{\partial^2 G^e}{\partial T \partial f}\right|_p,
\end{equation}
while the field-mechanical coupling is given by
\begin{equation}
\label{eq:mechfieldMP}
\xi_{T,N}=\left. \frac{\partial^2 G^e}{\partial p \partial f}\right|_T,
\end{equation}
and the field-chemical coupling is expressed by
\begin{equation}
\label{eq:fieldchemMP}
\chi_{T,p}=\left. \frac{\partial^2 G^e}{\partial T \partial N}\right|_T.
\end{equation}
Equation (\ref{eq:thefieldMP}) implies that for $\gamma_{p,N}=0$ the presence of the field does not affect the entropy of the system, whereas equations (\ref{eq:mechfieldMP}) and (\ref{eq:fieldchemMP}) imply that for $\xi_{T,N}=0$ or $\chi_{T,p}=0$ it is the volume or the chemical potential, respectively, to not be affected by the field. 

It is instructive to return briefly to the example of the protein mentioned in section 2. A protein tendency to fold is indicated by $S_{T,p,N}$ which measures the changes in the folding as hydrophobic interactions increase. The other material properties mentioned above will change depending on the configuration assumed by the protein and thus will depend on $f$. For the sake of clarity, we stress again that our approach concerns transformations among equilibrium states, while time dependences are not considered.

\section{Fluctuations of the thermodynamic state}

Small system fluctuations can be of the same order as the average value (e.g., energy fluxes between system and environment can be of the same order as the internal energy). As a result, the system is hardly ever observed in its average thermodynamic state, but typically resides in a generic fluctuating condition. As with the well-known theory of thermodynamic fluctuations \cite{landau1969statistical,callen2006thermodynamics}, this does not imply that the system is out of equilibrium, the latter defined with respect to the average, which here is assumed not to change. As the system size is increased, these fluctuations become undetectable and the thermodynamics reduces to the description of the average.

\subsection{Small system ensembles}
\label{sec:fluct}
For a given thermodynamic state $\hat{U}, \hat{V}, \hat{N}$, and $\hat{L}$, where $\hat{\cdot}$ denotes the random variables, let us define the fluctuating entropy of the system as the number of microscopic configurations through the Boltzmann postulate,
\begin{equation}
\label{eq:boltzentr}
\hat{S}(\hat{U},\hat{V},\hat{N},\hat{L})=k\ln \Omega(\hat{U},\hat{V},\hat{N},\hat{L}),
\end{equation}
where $k$ is the Boltzmann constant and $\Omega$ is the number of microstates associated to the given macrostate. The total differential of (\ref{eq:boltzentr}) multiplied by $T$ yields equation (\ref{eq:gibbseq}). Depending on the constraints set on the interaction between system and environment, different ensembles describe the thermodynamic fluctuations about the average state. These ensembles are obtained from the fundamental postulate of equal probability of the microstates and the definition of entropy (\ref{eq:boltzentr}). For example, when the extensive variables $U$, $V$, and $L$ fluctuate through the contact with an environment at constant intensive variables $T$, $p$, and $f$, the probability density function (pdf) for the fluctuations of $\hat{U}$, $\hat{V}$, and $\hat{L}$ is 
\begin{equation}
	\label{eq:pTfens}
	p_{pTf}(\hat{U},\hat{V},\hat{L}) = \Omega_0 e^{\beta (G^e-\hat{U}-p\hat{V}-f\hat{L}+T \hat{S})},
\end{equation}
where $\Omega_0$ is a normalization constant, $\beta=1/kT$, and the subscript in $p_{pTf}$ indicates the variables associated to the environment, which are constant. The ensemble (\ref{eq:pTfens}) has been used to describe the behavior of a macromolecule immersed in a solution \cite{jarzynski2017stochastic}, where the term $f \hat{L}$ was referred to as the solvation Hamiltonian of the mean force, although this was not expressed as a conjugate pair.

Regardless of which variable fluctuates, the general form of the pdf of an ensemble is proportional to the exponential of $\beta$ multiplied by a free energy difference: the free energy of the average state $\phi$ ($\phi=G^e$ in equation (\ref{eq:pTfens})) and a fluctuating free energy  $\hat{\phi}$ ($\hat{\phi}=\hat{G^e}=\hat{U}+p\hat{V}+f\hat{L}-T \hat{S}$ in equation (\ref{eq:pTfens})). The specific free energy appearing in the exponent is the Legendre Transform of $U$ with respect to the fluctuating variables (e.g., for $G^e$ the Laplace transform is done with respect to the fluctuating variables $U$, $V$, and $L$). For a generic ensemble of fluctuating variables $\hat{\bf X}$, in contact with an environment at constant intensive variables ${\bf y}$, we thus can write a general expression for the pdf of the thermodynamic state as \cite{landau1969statistical,callen2006thermodynamics,peliti2011statistical}
\begin{equation}
\label{eq:ens1}
p_\textbf{y}({\bf \hat{X}}) =\Omega_0 e^{\beta (\phi({\bf y})-\hat{\phi}({\bf y},\hat{\bf X}))}
\end{equation}
where $\phi({\bf y})$ is the free energy of the average thermodynamic state, $\hat{\phi}({\bf y},\hat{\bf X})$ is the fluctuating free energy, and ${\bf y}$ is the vector of intensive variables associated to the fluctuating variables $\hat{\bf X}$. Through the inclusion of the additional degree of freedom $L$ in the list of ${\bf X}$ in (\ref{eq:boltzentr}), equation (\ref{eq:ens1}) generalizes the theory of thermodynamic fluctuations \cite{landau1969statistical,callen2006thermodynamics} to small systems.

\subsection{The unconstrained ensemble}

Hill's unconstrained ensemble \cite{hill2001different} is one among the novel thermodynamic ensembles, described above, that can be realized due to the additional degree of freedom. Originally introduced to describe colloidal clusters in a solvent \cite{hill1963thermodynamics}, its use has been extended to metastable states of liquid cluster in a supersaturated gaseous phase \cite{hill1998extension} and to systems with long-range interaction \cite{latella2017long}. Such an ensemble corresponds to a completely open system with fluctuating energy, volume, and number of particles in contact with a reservoir at constant $T$, $p$, and $\mu$. The potential associated to it is obtained by Legendre transform of the internal energy with respect to $S$, $V$, and $N$, $U[S,V,N]=E=U+pV-\mu N-T S=fL$, and its differential $dE=-SdT+Vdp-Nd\mu$ (here $L$ is constant, so $fdL=0$). It is thus straightforward to see that this ensemble is not realizable in macroscopic systems because $\mu$, $T$, and $p$ are related by the familiar Gibbs-Duhem relation, $Nd\mu=-SdT+Vdp$, and the Legendre transform of $U$ yields $U[S,V,N]=0$. From the definition of $E$ and from equation (\ref{eq:ens1}), the pdf associated to the unconstrained unsemble is 
\begin{equation}
\label{eq:uncostens}
p_{pT\mu}(\hat{U},\hat{V},\hat{N}) = \Omega_0 e^{\beta (E-\hat{U}-p\hat{V}+\mu \hat{N}+T \hat{S})}.
\end{equation}


\subsection{Thermodynamics of the fluctuating state}

Having described the thermodynamics of the average thermodynamic state (section \ref{sec:smallsyst}) and the statistics of the fluctuations about it (section \ref{sec:fluct}), we now turn to the derivation of the thermodynamics of the fluctuating state. The free energy difference in the exponent of (\ref{eq:ens1}) corresponds, from a physical point of view, to an availability \cite{kestin1979course},
\begin{equation}
\label{eq:avail}
\hat{R}(\hat{\textbf{X}})=\hat{\phi}-\phi,
\end{equation}
which represents the maximum work extractable from the system while undergoing a fluctuation. Equivalently, the availability (\ref{eq:avail}) is the minimum work required to isolate the system from the environment \cite{dellesite2017partitioning} in a generic state $\hat{\textbf{X}}$ different from the average thermodynamic state. Hence equation (\ref{eq:avail}) measures the distance, in energetic terms, of the fluctuation from the average.



To the availability $\hat{R}$ there correspond entropic forces that counteract the fluctuation. To derive these forces, it is necessary to identify the proper entropy function. Substituting in (\ref{eq:avail}) the definition of fluctuating free energy, $\hat{\phi}({\bf y},\hat{\bf X})=T(\textbf{Y} \hat{\textbf{X}}-\hat{S}(\hat{\textbf{X}}))$, the availability can be written as
\begin{equation}
\label{eq:avail2}
\hat{R}=(T\textbf{Y} \hat{\textbf{X}}-\phi)-T\hat{S}(\hat{\textbf{X}}),
\end{equation}
where $\textbf{Y}$ are the intensive variables (their average value) conjugate to $\hat{\textbf{X}}$ in entropy representation (e.g., $1/T$, $p/T$). Introducing
\begin{equation}
\label{eq:stochentr}
\hat{\mathcal{S}}=-\phi/T+\textbf{Y} \hat{\textbf{X}},
\end{equation}
we can re-express $\hat{R}$ as the difference
\begin{equation}
\label{eq:avail3}
\hat{R}=T(\hat{\mathcal{S}}-\hat{S}),
\end{equation}
between the entropies (\ref{eq:stochentr}) and (\ref{eq:boltzentr}), and since the availability is always positive, $\hat{\mathcal{S}}\ge\hat{S}$. The entropy $\hat{\mathcal{S}}$ has been used as a non-equilibrium entropy in stochastic thermodynamics (e.g., \cite{seifert2012stochastic}). Specifically, in this context the fluctuating entropy is defined as the negative logarithm of the probability of occurrence of a microstate, $p_\textbf{y}(\hat{z})$, where $\hat{z}$ represents the phase space coordinates. From the well-known relationship between the probability of a microstate and the probability of a macrostate \cite{greene1951formalism}, $p_\textbf{y}(\hat{z}) \Omega d\hat{\bf X}=p_\textbf{y}({\bf \hat{\bf X}})d\hat{X}$, one obtains
\begin{equation}
\label{eq:pmicro}
p_\textbf{y}(\hat{\bf z})=e^{\frac{1}{T} (\frac{\phi(\bf y)}{T}-{\bf Y}\cdot {\bf X}(\hat{z}))},
\end{equation}
so that computing its logarithm yields $-k \ln (p_\textbf{y}(\hat{z}))=-\phi/T+{\bf Y}{\bf X}$, which corresponds to $\mathcal{S}$ in equation (\ref{eq:stochentr}).

It is logical to wonder what thermodynamic formalism emerges from the two entropies above and whether they are appropriate to describe the thermodynamics of fluctuations. To this aim, it is important to consider that, since the values of $\textbf{Y}$ are kept constant, the entropy $\hat{\mathcal{S}}$ is a linear function of $\hat{\textbf{X}}$ (see Figure \ref{Fig:figure1}, panels (a) and (b)). Equation (\ref{eq:stochentr}), in fact, is not the result of a total differential, as $\hat{\mathcal{S}}$ is not a linear homogeneous function of $\hat{\textbf{X}}$. Because this basic property is missing, there is no fundamental equation of the form $\hat{\mathcal{S}}=\hat{\mathcal{S}}(\hat{\textbf{X}})$ that defines the thermodynamic state of the system during a generic fluctuation. On the contrary, the entropy $\hat{S}$ defined by equation (\ref{eq:boltzentr}) is a concave function (Figure \ref{Fig:figure1}(b)) and its total differential yields the Euler relation
\begin{equation}
\label{eq:stochentr3}
\hat{S}=-\hat{\phi}/\hat{T}+\hat{\textbf{Y}} \hat{\textbf{X}},
\end{equation}
where $\hat{\textbf{Y}}=\partial \hat{S}/\partial \hat{\textbf{X}}$. This is because $\hat{\mathcal{S}}$ in (\ref{eq:boltzentr}) is linear and $\hat{R}$ grows with the size of the fluctuation, namely with the distance of ${\bf \hat{X}}$ from its average value ${\bf X}$. As a result, $\hat{S}$ is a linear homogeneous function and is suitable to describe the thermodynamics of the generic fluctuating state. 

Equation (\ref{eq:stochentr3}) also shows that along with the extensive variables, also the intensive variables $\hat{\textbf{Y}}$ are fluctuating quantities. When the system is away from the average, the entropic force that opposes these fluctuations can be computed as the difference between the fluctuating and the average intensive quantities,
\begin{equation}
\label{eq:entrforce}
\hat{\mathcal{F}}=\nabla_\textbf{X} (S-\hat{S})=\textbf{Y}-\hat{\textbf{Y}}.
\end{equation}
The quantity in (\ref{eq:entrforce}) can be positive or negative, and goes to zero only when the system is in equilibrium with its environment, namely whenever the thermodynamic state passes by its average state, $\hat{\textbf{X}}=\textbf{X}$. Under this condition, one also has $\hat{\phi}=\phi$ and $\hat{\textbf{Y}}=\textbf{Y}$, so that the two entropies coincide, $\hat{\mathcal{S}}=\hat{S}$, and of course $\hat{R}=0$. Equation (\ref{eq:stochentr3}) can then be considered as the logical extension of equation (\ref{eq:euler}) to small systems fluctuating because of the interaction with the environment. 

\subsection{Equivalence of ensembles in small systems}

Without including $L$ as an additional degree of freedom, the classical thermodynamic framework when applied to small systems is non-additive in the presence of interactions with the environment and gives rise to nonconcave entropy envelopes. As discussed in detailed by Touchette [2015], this in turn makes the thermodynamic ensembles non-equivalent. With the inclusion of the extensive variable $L$, as presented here, the entropy $S$ is concave, thus guaranteeing that any couple of ensembles is equivalent. 

As an example, it can be readily shown that the extended canonical ensemble ($NTf$-ensemble) corresponds to the $pTf$-ensemble, conditional on a given volume. The marginal pdf $p_{pTf}(V)$ is
\begin{equation}
p_{pTf}(\hat{V})=\int p_{pTf}(\hat{U},\hat{V},\hat{L})d\hat{U}d\hat{L}=\Omega_0 e^{\beta (G^e-p\hat{V})} \int e^{-\beta(\hat{U}+f\hat{L}-T\hat{S})} d\hat{U}d\hat{L},
\end{equation}
and in turn the conditional pdf
\begin{equation}
p_{pTf}(\hat{U},\hat{L}|\hat{V}=V)=\frac{\Omega_0 e^{\beta (G^e-\hat{U}-pV-f\hat{L}+T \hat{S})}}{\Omega_0 e^{\beta (G^e-pV)} \int e^{-\beta(\hat{U}+f\hat{L}-T\hat{S})} d\hat{U}d\hat{L}}=\frac{e^{\beta (-\hat{U}-f\hat{L}+T \hat{S})}}{\int e^{-\beta(\hat{U}+f\hat{L}-T\hat{S})} d\hat{U}d\hat{L}},
\end{equation}
which, substituting $\Omega_0 (V) e^{\beta F^e}=\int e^{\beta(\hat{U}+f\hat{L}-T\hat{S})} d\hat{U}d\hat{L}$, yields
\begin{equation}
\label{eq:enscond}
p_{pTf}(\hat{U},\hat{L}|\hat{V}=V)=\Omega_0'(V)e^{\beta (F^e-\hat{U}-f\hat{L}+T \hat{S})},
\end{equation}
where because $S$ is a function also of $V$, the normalization constant depends on $V$. It can be clearly seen that (\ref{eq:enscond}) is nothing but the extended canonical ensemble, derivable also from equation (\ref{eq:ens1}). Analogous considerations can be made for any other pair of ensembles.

Failing to consider $L$ a random variable in conditions in which its values are not constant, as would be the case of a typical $pT$-ensemble, clearly corresponds to using the incorrect ensemble pdf and thus yields inaccurate statistics of $\hat{{\bf X}}$.

\section{Small system of ideal gas}

When the small system consists of a fraction $N/M$ of tagged particles of an ideal gas made of $M$ particles, there is no additional degree of freedom involved and, in turn, the thermodynamic description of the system reduces to the classical one. The characteristic aspect of ideal gases is that, since particles interact at practically infinitesimally small distances during correspondingly short time intervals, their particles interaction energy is negligible compared to their total energy, while their kinetic energy is rather large. Their potential energy, although fundamental for the particles to settle on a thermodynamic state, is in fact often taken to vanish. In turn, also the interaction energy of its particles with the environment (the rest of the gas) is negligible compared to their total energy. This holds for one particle as well as for any size of the system. Therefore, in terms of the thermodynamic equations (\ref{eq:gibbseq}) and (\ref{eq:euler}), we may state that both $f$ and $L$ practically vanish, so that (similarly to the case of a macroscopic system) the internal energy of a small ideal gas system is a function of only $S$, $V$, and $N$. 

The small ideal gas system still undergoes fluctuations of large relative size and a thermodynamics of the average and of the fluctuating state can be formulated as described above. However, in this case the thermodynamic degrees of freedom do not include $L$. In addition, the negligible mutual interaction between the particles and between system and environment is reflected into the pdf of the fluctuations. Considering the canonical ensemble, for example, the pdf for the energy fluctuations, obtained by substituting in (\ref{eq:ens1}) the expressions for the entropy and Helmholtz free energy of the ideal gas (Sackur-Tetrode equation \cite{garrod1995statistical}), has the form of a Gamma distribution, $p_T(\hat{U}) \propto  \hat{U}^{N/2}\beta^{3/2N}e^{-\beta \hat{U}}$. Such a pdf has the property of being factorizable \cite{callen2006thermodynamics}. In other words, the internal energy $U$ is additive for any value of $N$ and its pdf for a system of $N$ particles is the same as the one of the individual constituents. Thus, any system with negligible interaction between its constituents and with the environment, not necessarily of ideal gas, retains the additivity of its internal energy and, in turn, the factorizability of the ensemble pdf. 
 

\section{Single particle interacting with the environment}
\label{sec:app}
\begin{figure}
	\includegraphics[width=13 cm]{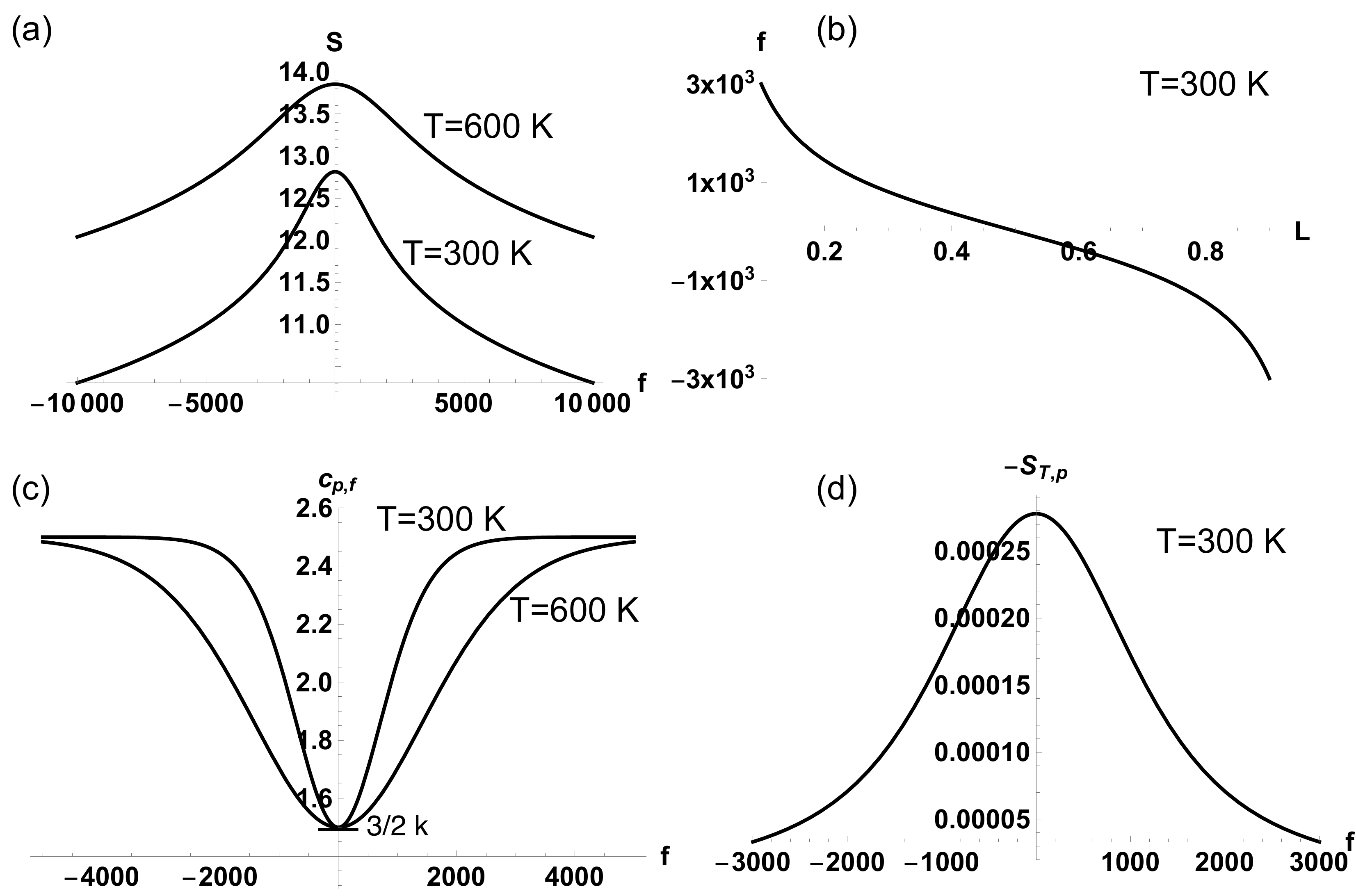}
	\centering
	\caption{\footnotesize (a) Entropy of the small system of section 4 as a function of the $f$, showing the ordering effect caused by the interaction. (b) Equation of state $f(U,L)$ plotted as a function of $L$. (c) Isobaric heat capacity as a function of $f$. The graph highlights the increase in heat capacity due to the field and its dependence on temperature. (d) Susceptibility as a function of $f$. Graphs were drawn for $L_0=1$ and $k=1$.}
	\label{Fig:figure1}
\end{figure}

In light of the new framework introduced above, we revisit the typical example used in stochastic thermodynamics, a single particle interacting with the environment. Consider a particle immersed in a fluid confined in a volume $V$ serving as a heat bath at temperature $T$. Let the particle be also subject to the influence of a force field arising from its interaction with the external environment. For example, this force may be generated by an optical tweezer manipulating the system. We indicate with $l$ the position of the particle within the volume $V$ along the direction of the interaction force of modulus $f$. We assume $f$ to be constant for simplicity, but other assumptions would not affect our conclusions. The Hamiltonian is $H=1/2 m\textbf{v}^2+f l$, so that the probability density function of the microstate (i.e., velocity and position) is
\begin{equation}
p(\hat{\bf z})=p(\hat{\bf v},\hat{l})=\frac{1}{Z}e^{-\beta (\frac{m \hat{\bf v}^2}{2}+f\hat{l})}.
\end{equation}
From the partition function $Z=e^{-\beta F^e}=\int\limits_{0}^{L_{0}} \int\limits_{-\infty}^{\infty} e^{-\beta (\frac{m \hat{\bf v}^2}{2}+f\hat{l})}d^3\hat{v} \cdot d\hat{l}$, where $L_0$ is the length of the confining volume along the direction of the force $f$ and $F^e=U+fL-TS$ is the extended Helmholtz free energy, all the thermodynamic quantities can be derived. The entropy reads
\begin{equation}
\label{eq:avethermoquant}
S=-\frac{\partial F^e}{\partial T}=k \log \left(\frac{2 \sqrt{2} \pi ^{3/2} \psi }{f \left(\frac{1}{k T}\right)^{5/2}}\right)+\frac{f L_0}{T \psi }+\frac{5 k}{2},
\end{equation}
where $\psi=1-e^{\frac{f L_0}{k T}}$. The ensemble average of $\hat{l}$ is
\begin{equation}
\label{eq:avethermoquant2}
L=\frac{\partial F^e}{\partial f}=\frac{k T}{f}-\frac{L_0}{e^{\frac{f L_0}{k T}}-1},
\end{equation}
and in turn the internal energy is
\begin{equation}
\label{eq:intener}
U=F^e-fL+ST=\frac{3}{2}kT.
\end{equation}

Combining the entropy (\ref{eq:avethermoquant}) with equations (\ref{eq:avethermoquant2}) and (\ref{eq:intener}) provides the fundamental equation of the single particle in the heat bath, $S=S(U,L)$. The latter is a monotonically increasing function of the energy $\hat{U}$; however, when plotted as a function of $\hat{L}$, it has a maximum at $f=0$ and decreases as $f$ increases (Figure \ref{Fig:figure1}(a)). This symmetrical bell shape originates from the ordering effect brought about by the interaction with the environment, which forces the particle to spend more time in some positions rather than others. While in the absence of the field the particle freely explores the whole volume and the entropy is maximum, the particle is pushed towards the wall when the interaction is turned on (Figure \ref{Fig:figure1}(b)). 

The interaction with the environment also affects the material properties of the system. From equations (\ref{eq:matprop}), the isobaric heat capacity is $c_{p,f}=3/2 k$ at $f=0$ and increases as the interaction becomes stronger. An important result is that the heat capacity becomes a decreasing function of temperature in the presence of the interaction (Figure \ref{Fig:figure1}(c)). The susceptibility is a negative quantity in our convention, as a positive interaction (i.e., attractive) reduces the distance from the wall, while its modulus decreases with $f$. Given the presence of thermal agitation, it is harder for the environment to move the average position of the particle $L$ closer to a wall (Figure \ref{Fig:figure1}(d)).
\begin{figure}
	\includegraphics[width=13 cm]{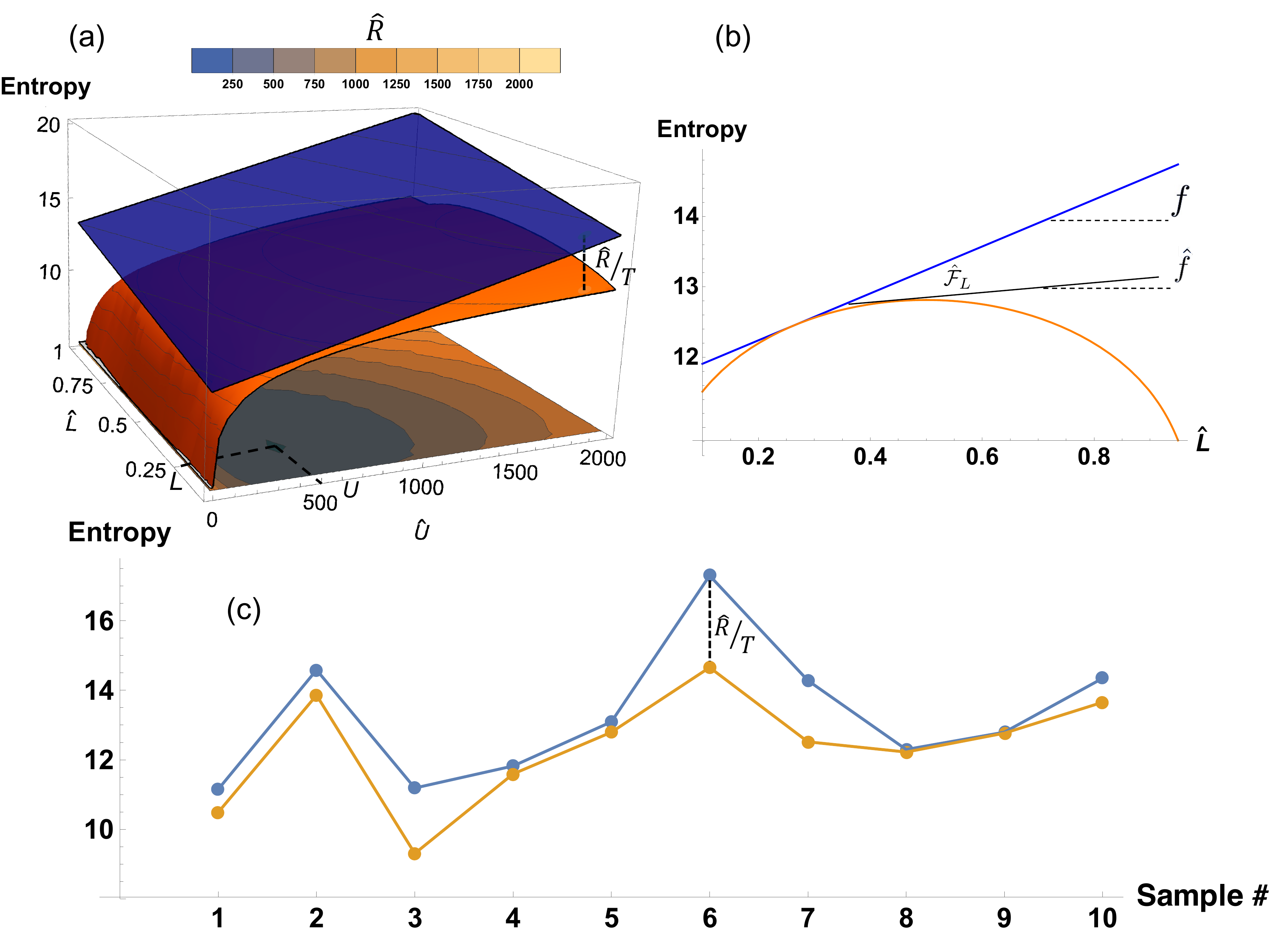}
	\centering
	\caption{\footnotesize (a) Entropies $\mathcal{S}$ (blue) and $S$ (orange) as a function of the extensive variables $\hat{U}$ and $\hat{L}$. Contour plot of the availability $R$, on the bottom $\hat{U}$-$\hat{L}$ space, measuring the distance between the two entropies (equation (\ref{eq:avail4})). (b) Entropies $\mathcal{S}$ (blue) and $S$ (orange) as a function of $\hat{L}$ for internal energy equal to its average value, $\hat{U}=U$. (c) Entropies $\mathcal{S}$ (blue) and $S$ (orange) from Gibbs sampling of the thermodynamic state from the ensemble pdf (\ref{eq:pTfens2}).}
	\label{Fig:figure2}
\end{figure}

The thermodynamic fluctuations of $U$ and $L$ are described by the $NTf$-ensemble, namely an extended canonical ensemble with the form
\begin{equation}
\label{eq:pTfens2}
p_{NTf}(\hat{U},\hat{L}) = \Omega_0 e^{\beta (F^e-\hat{U}-f\hat{L}+T \hat{S})}.
\end{equation}
As the system fluctuates between thermodynamic states (Figure \ref{Fig:figure2}(c)), the thermodynamics of the system can be described, according to (\ref{eq:stochentr3}), as
\begin{equation}
\hat{F}^e=\hat{U}+\hat{f}\hat{L}-\hat{T}\hat{S},
\end{equation}
where the entropy $\hat{S}$ is computed from the fundamental equation $\hat{S}=\hat{S}(\hat{U},\hat{L})$. In contrast, the entropy $\hat{\mathcal{S}}$ is computed from (\ref{eq:stochentr}). The two entropies are plotted as a function of $\hat{U}$ and $\hat{L}$ in Figure \ref{Fig:figure2}(a), where it becomes clear that, while $\hat{S}$ is a concave function of its argument, the entropy $\hat{\mathcal{S}}$ is linear in both $\hat{U}$ and $\hat{L}$, resulting in a plane that is tangent to the entropy $\hat{S}$ at the average state, i.e., $\hat{U}=U$ and $\hat{L}=L$.

When the fluctuating state is away from the average, the two entropies diverge, giving rise to an availability
\begin{equation}
\label{eq:avail4}
\hat{R}=\hat{F}^e-F^e=T(\hat{\mathcal{S}}-\hat{S}),
\end{equation}
which is shown in the contour plot in Figure \ref{Fig:figure2}(a) as a function of $\hat{U}$ and $\hat{L}$. The availability is zero at $\hat{U}$ and $\hat{L}$ equal to the average values, and it increases with the distance from the average. At the same time, entropic forces emerge,
\begin{equation}
\hat{\mathcal{F}}_U=\frac{1}{T}-\frac{1}{\hat{T}} \quad \text{and} \quad \hat{\mathcal{F}}_L=\frac{f}{T}-\frac{\hat{f}}{\hat{T}},
\end{equation}
that tend to bring the fluctuating state back to its average. We emphasize again that, whether or not there is an interaction $f$ with the environment (e.g., $f=0$ for an ideal gas), the fluctuations that originate from the contact with the environment are spontaneous. That is, the system remains in thermodynamic equilibrium with a constant average state. As the size of the system increases, these fluctuations become negligible relative to the mean and the thermodynamics of the system reduces to a description of the average state. 

\section{Conclusions}

We constructed a thermodynamic formalism for small systems by explicitly incorporating the interaction with the environment through conjugate intensive and extensive variables (Section 2). This gave rise to an extended thermodynamic space with novel thermodynamic potentials and material properties (i.e., susceptibility) that quantify the effect of the interaction on the small system.  

Because fluctuations in small systems are large and likely, the thermodynamics can be extended to a description of the fluctuations and of the generic fluctuating state (Section 3). The proper definition of entropy yields the Euler relation for the fluctuating state and the availability, which appears in the exponent of the ensemble pdf, measures in energetic terms the distance from the average state. The present formalism encompasses and extends both the nano-thermodynamics by Hill \cite{hill1963thermodynamics,hill2001different} and the statistical mechanics of systems with strong coupling \cite{seifert2016strongcoupling,jarzynski2017stochastic}, see equation (\ref{eq:pTfens}). Furthermore, it was shown that the thermodynamic ensembles are all equivalent. 

We discussed in detail the case of small system that is a portion of a macroscopic ideal gas system (Section 4). In this case, because the potential energy within the particles is negligible, the thermodynamics of small systems becomes equivalent to the corresponding macroscopic thermodynamics (i.e, there is no additional degree of freedom) and, in turn, the ensemble pdf describing the energy fluctuations (these are still large) is factorizable. As an illustration of our framework, we revisited the example of a particle immersed in a heat bath, under the influence of an external force representing the interaction with the external environment.   

Application of classical thermodynamics to small systems or systems with long-range interactions typically gives rise to non-concave envelops of the fundamental equation \cite{campa2009statistical} and, as a consequence, their ensembles are inequivalent. Here we showed that ensemble equivalence is restored within the presented formalism, which considers explicitly the interaction with the environment. With some caution, these considerations on small systems apply also to systems with long-range interactions \cite{latella2017long}. Being defined by the fact that the two particle potential decays over long distances compared to the volume of the system, systems with long-range interactions always interact with their environment, so that in this sense they can be considered analogous to small systems.


\begin{thebibliography}{}
	
	\bibitem[Abendroth et~al., 2015]{abendroth2015controlling}
	Abendroth, J.~M., Bushuyev, O.~S., Weiss, P.~S., and Barrett, C.~J. (2015).
	\newblock Controlling motion at the nanoscale: rise of the molecular machines.
	\newblock {\em ACS nano}, 9(8):7746--7768.
	
	\bibitem[An et~al., 2015]{an2015experimental}
	An, S., Zhang, J.-N., Um, M., Lv, D., Lu, Y., Zhang, J., Yin, Z.-Q., Quan, H.,
	and Kim, K. (2015).
	\newblock Experimental test of the quantum jarzynski equality with a
	trapped-ion system.
	\newblock {\em Nature Physics}, 11(2):193.
	
	\bibitem[Bedeaux and Kjelstrup, 2018]{bedeaux2018hill}
	Bedeaux, D. and Kjelstrup, S. (2018).
	\newblock Hill's nano-thermodynamics is equivalent with gibbs' thermodynamics
	for curved surfaces.
	\newblock {\em arXiv preprint arXiv:1806.03444}.
	
	\bibitem[Bustamante et~al., 2004]{bustamante2004mechanical}
	Bustamante, C., Chemla, Y.~R., Forde, N.~R., and Izhaky, D. (2004).
	\newblock Mechanical processes in biochemistry.
	\newblock {\em Annual review of biochemistry}, 73(1):705--748.
	
	\bibitem[Calabrese and Porporato, 2019]{calabrese2019origin}
	Calabrese, S. and Porporato, A. (2019).
	\newblock Origin of negative temperatures in systems interacting with external
	fields.
	\newblock {\em Physics Letters A}, 383(18):2153--2158.
	
	\bibitem[Callen, 2006]{callen2006thermodynamics}
	Callen, H. (2006).
	\newblock {\em Thermodynamics and an introduction to thermodynamics}.
	\newblock Student Edition. Wiley India Pvt. Limited.
	
	\bibitem[Campa et~al., 2009]{campa2009statistical}
	Campa, A., Dauxois, T., and Ruffo, S. (2009).
	\newblock Statistical mechanics and dynamics of solvable models with long-range
	interactions.
	\newblock {\em Physics Reports}, 480(3):57--159.
	
	\bibitem[Chamberlin, 2014]{chamberlin2014big}
	Chamberlin, R. (2014).
	\newblock The big world of nanothermodynamics.
	\newblock {\em Entropy}, 17(1):52--73.
	
	\bibitem[Chamberlin, 2000]{chamberlin2000mean}
	Chamberlin, R.~V. (2000).
	\newblock Mean-field cluster model for the critical behaviour of ferromagnets.
	\newblock {\em Nature}, 408(6810):337.
	
	\bibitem[Ciliberto et~al., 2010]{ciliberto2010fluctuations}
	Ciliberto, S., Joubaud, S., and Petrosyan, A. (2010).
	\newblock Fluctuations in out-of-equilibrium systems: from theory to
	experiment.
	\newblock {\em Journal of Statistical Mechanics: Theory and Experiment},
	2010(12):P12003.
	
	\bibitem[Collin et~al., 2005]{collin2005verification}
	Collin, D., Ritort, F., Jarzynski, C., Smith, S.~B., Tinoco~Jr, I., and
	Bustamante, C. (2005).
	\newblock Verification of the crooks fluctuation theorem and recovery of rna
	folding free energies.
	\newblock {\em Nature}, 437(7056):231.
	
	\bibitem[Corti and Soto-Campos, 1998]{corti1998deriving}
	Corti, D.~S. and Soto-Campos, G. (1998).
	\newblock Deriving the isothermal--isobaric ensemble: The requirement of a
	“shell” molecule and applicability to small systems.
	\newblock {\em The Journal of chemical physics}, 108(19):7959--7966.
	
	\bibitem[Dobson, 2003]{dobson2003protein}
	Dobson, C.~M. (2003).
	\newblock Protein folding and misfolding.
	\newblock {\em Nature}, 426(6968):884.
	
	\bibitem[Esposito et~al., 2010]{esposito2010entropy}
	Esposito, M., Lindenberg, K., and Van~den Broeck, C. (2010).
	\newblock Entropy production as correlation between system and reservoir.
	\newblock {\em New Journal of Physics}, 12(1):013013.
	
	\bibitem[Garrod, 1995]{garrod1995statistical}
	Garrod, C. (1995).
	\newblock {\em Statistical mechanics and thermodynamics}.
	\newblock Oxford University Press, USA.
	
	\bibitem[Gibbs, 1928]{gibbs1928collected}
	Gibbs, J.~W. (1928).
	\newblock {\em The collected works of J. Willard Gibbs, volume I:
		thermodynamics}.
	\newblock Yale University Press, New Haven, CT, USA.
	
	\bibitem[Goldstein et~al., 2017]{goldstein2017statistical}
	Goldstein, S., Huse, D.~A., Lebowitz, J.~L., and Sartori, P. (2017).
	\newblock Statistical mechanics and thermodynamics of large and small systems.
	\newblock {\em arXiv preprint arXiv:1712.08961}.
	
	\bibitem[Greene and Callen, 1951]{greene1951formalism}
	Greene, R.~F. and Callen, H.~B. (1951).
	\newblock On the formalism of thermodynamic fluctuation theory.
	\newblock {\em Physical Review}, 83(6):1231.
	
	\bibitem[Hilbert et~al., 2014]{hilbert2014thermodynamic}
	Hilbert, S., H{\"a}nggi, P., and Dunkel, J. (2014).
	\newblock Thermodynamic laws in isolated systems.
	\newblock {\em Physical Review E}, 90(6):062116.
	
	\bibitem[Hill, 1963]{hill1963thermodynamics}
	Hill, T.~L. (1963).
	\newblock {\em Thermodynamics of small systems}.
	\newblock Courier Corporation.
	
	\bibitem[Hill, 2001]{hill2001different}
	Hill, T.~L. (2001).
	\newblock A different approach to nanothermodynamics.
	\newblock {\em Nano Letters}, 1(5):273--275.
	
	\bibitem[Hill and Chamberlin, 1998]{hill1998extension}
	Hill, T.~L. and Chamberlin, R.~V. (1998).
	\newblock Extension of the thermodynamics of small systems to open metastable
	states: An example.
	\newblock {\em Proceedings of the National Academy of Sciences},
	95(22):12779--12782.
	
	\bibitem[Horowitz and Jarzynski, 2008]{horowitz2008comment}
	Horowitz, J. and Jarzynski, C. (2008).
	\newblock Comment on “failure of the work-hamiltonian connection for
	free-energy calculations”.
	\newblock {\em Physical review letters}, 101(9):098901.
	
	\bibitem[Hummer and Szabo, 2001]{hummer2001free}
	Hummer, G. and Szabo, A. (2001).
	\newblock Free energy reconstruction from nonequilibrium single-molecule
	pulling experiments.
	\newblock {\em Proceedings of the National Academy of Sciences},
	98(7):3658--3661.
	
	\bibitem[Hunjan and Eu, 2010]{hunjan2010voronoi}
	Hunjan, J.~S. and Eu, B.~C. (2010).
	\newblock The voronoi volume and molecular representation of molar volume:
	equilibrium simple fluids.
	\newblock {\em The Journal of chemical physics}, 132(13):134510.
	
	\bibitem[Jarzynski, 2017]{jarzynski2017stochastic}
	Jarzynski, C. (2017).
	\newblock Stochastic and macroscopic thermodynamics of strongly coupled
	systems.
	\newblock {\em Physical Review X}, 7(1):011008.
	
	\bibitem[Kauzmann, 1959]{kauzmann1959some}
	Kauzmann, W. (1959).
	\newblock Some factors in the interpretation of protein denaturation.
	\newblock In {\em Advances in protein chemistry}, volume~14, pages 1--63.
	Elsevier.
	
	\bibitem[Kestin, 1979]{kestin1979course}
	Kestin, J. (1979).
	\newblock {\em A Course In Thermodynamics}.
	\newblock Number v. 1 in A Course in Thermodynamics. Taylor \& Francis.
	
	\bibitem[Kinbara and Aida, 2005]{kinbara2005toward}
	Kinbara, K. and Aida, T. (2005).
	\newblock Toward intelligent molecular machines: directed motions of biological
	and artificial molecules and assemblies.
	\newblock {\em Chemical reviews}, 105(4):1377--1400.
	
	\bibitem[Kunz and Berry, 1994]{kunz1994multiple}
	Kunz, R.~E. and Berry, R.~S. (1994).
	\newblock Multiple phase coexistence in finite systems.
	\newblock {\em Physical Review E}, 49(3):1895.
	
	\bibitem[Landau and Lifshitz, 1969]{landau1969statistical}
	Landau, L.~D. and Lifshitz, E.~M. (1969).
	\newblock {\em Statistical Physics: V. 5: Course of Theoretical Physics}.
	\newblock Pergamon press.
	
	\bibitem[Latella et~al., 2017]{latella2017long}
	Latella, I., P{\'e}rez-Madrid, A., Campa, A., Casetti, L., and Ruffo, S.
	(2017).
	\newblock Long-range interacting systems in the unconstrained ensemble.
	\newblock {\em Physical Review E}, 95(1):012140.
	
	\bibitem[Miller and Anders, 2017]{miller2017entropy}
	Miller, H.~J. and Anders, J. (2017).
	\newblock Entropy production and time asymmetry in the presence of strong
	interactions.
	\newblock {\em Physical Review E}, 95(6):062123.
	
	\bibitem[Nelson, 2004]{nelson2004biological}
	Nelson, P. (2004).
	\newblock {\em Biological physics}.
	\newblock WH Freeman New York.
	
	\bibitem[Peliti, 2008]{peliti2008comment}
	Peliti, L. (2008).
	\newblock Comment on “failure of the work-hamiltonian connection for
	free-energy calculations”.
	\newblock {\em Physical review letters}, 101(9):098903.
	
	\bibitem[Peliti, 2011]{peliti2011statistical}
	Peliti, L. (2011).
	\newblock {\em Statistical mechanics in a nutshell}.
	\newblock Princeton University Press.
	
	\bibitem[Qian, 2012]{qian2012hill}
	Qian, H. (2012).
	\newblock Hill’s small systems nanothermodynamics: a simple macromolecular
	partition problem with a statistical perspective.
	\newblock {\em Journal of biological physics}, 38(2):201--207.
	
	\bibitem[Schnell et~al., 2011]{schnell2011thermodynamics}
	Schnell, S.~K., Vlugt, T.~J., Simon, J.-M., Bedeaux, D., and Kjelstrup, S.
	(2011).
	\newblock Thermodynamics of a small system in a $\mu$t reservoir.
	\newblock {\em Chemical Physics Letters}, 504(4-6):199--201.
	
	\bibitem[Schnell et~al., 2012]{schnell2012thermodynamics}
	Schnell, S.~K., Vlugt, T.~J., Simon, J.-M., Bedeaux, D., and Kjelstrup, S.
	(2012).
	\newblock Thermodynamics of small systems embedded in a reservoir: a detailed
	analysis of finite size effects.
	\newblock {\em Molecular Physics}, 110(11-12):1069--1079.
	
	\bibitem[Seifert, 2012]{seifert2012stochastic}
	Seifert, U. (2012).
	\newblock Stochastic thermodynamics, fluctuation theorems and molecular
	machines.
	\newblock {\em Reports on Progress in Physics}, 75(12):126001.
	
	\bibitem[Seifert, 2016]{seifert2016strongcoupling}
	Seifert, U. (2016).
	\newblock First and second law of thermodynamics at strong coupling.
	\newblock {\em Phys. Rev. Lett.}, 116:020601.
	
	\bibitem[Sekimoto, 1998]{sekimoto1998langevin}
	Sekimoto, K. (1998).
	\newblock Langevin equation and thermodynamics.
	\newblock {\em Progress of Theoretical Physics Supplement}, 130:17--27.
	
	\bibitem[Shipley, 2010]{shipley2010plant}
	Shipley, B. (2010).
	\newblock {\em From plant traits to vegetation structure: chance and selection
		in the assembly of ecological communities}.
	\newblock Cambridge University Press.
	
	\bibitem[Site et~al., 2017]{dellesite2017partitioning}
	Site, L.~D., Ciccotti, G., and Hartmann, C. (2017).
	\newblock Partitioning a macroscopic system into independent subsystems.
	\newblock {\em Journal of Statistical Mechanics: Theory and Experiment},
	2017(8):083201.
	
	\bibitem[Talkner and H{\"a}nggi, 2016]{talkner2016open}
	Talkner, P. and H{\"a}nggi, P. (2016).
	\newblock Open system trajectories specify fluctuating work but not heat.
	\newblock {\em Physical Review E}, 94(2):022143.
	
	\bibitem[Uline and Corti, 2013]{uline2013molecular}
	Uline, M. and Corti, D. (2013).
	\newblock Molecular dynamics at constant pressure: allowing the system to
	control volume fluctuations via a “shell” particle.
	\newblock {\em Entropy}, 15(9):3941--3969.
	
	\bibitem[Van~den Broeck and Esposito, 2015]{van2015ensemble}
	Van~den Broeck, C. and Esposito, M. (2015).
	\newblock Ensemble and trajectory thermodynamics: A brief introduction.
	\newblock {\em Physica A: Statistical Mechanics and its Applications},
	418:6--16.
	
	\bibitem[Vilar and Rubi, 2008]{vilar2008failure}
	Vilar, J.~M. and Rubi, J.~M. (2008).
	\newblock Failure of the work-hamiltonian connection for free-energy
	calculations.
	\newblock {\em Physical review letters}, 100(2):020601.
	
	\bibitem[Zhang et~al., 2012]{zhang2012stochastic}
	Zhang, X.-J., Qian, H., and Qian, M. (2012).
	\newblock Stochastic theory of nonequilibrium steady states and its
	applications. part i.
	\newblock {\em Physics Reports}, 510(1-2):1--86.
	
\end{thebibliography}
\end{document}